\documentclass{PoS}
\pdfoutput=1
\newcommand{\be}{\begin{equation}}
\newcommand{\ee}{\end{equation}}
\newcommand{\bea}{\begin{eqnarray}}
\newcommand{\eea}{\end{eqnarray}}
\newcommand{\ba}{\begin{array}}
\newcommand{\ea}{\end{array}}

\newcommand{\pref}[1]{(\ref{#1})}

\newcommand{\mpit}{M_0^2}
\newcommand{\tct}{2c_2a^2}

\def\bqry{\begin{eqnarray}}
\def\eqry{\end{eqnarray}}

\def\lbl{\label}

\title{Pion Scattering in Wilson Chiral Perturbation Theory }

\ShortTitle{Pion Scattering in Wilson Chiral Perturbation Theory }

\author{Sinya Aoki\\
Graduate School of Pure and Applied Sciences\\
University of Tsukuba\\
Tsukuba, Ibaraki  305-8571\\
Japan\\
{\em and}\\
Riken BNL Research Center\\
Brookhaven National Laboratory\\
Upton, New York 11973\\
USA\\
E-mail: \email{saoki@het.ph.tsukuba.ac.jp}
}
\author{{Oliver B\"ar and \speaker{Benedikt Biedermann}}\\
                 Institute of Physics\\
        Humboldt University Berlin\\
	12489 Berlin\\
	Germany\\
	E-mail: \email{obaer@physik.hu-berlin.de,}\\
	\phantom{E-mail:} \email{benedikt.biedermann@physik.hu-berlin.de}
}

\abstract{
We compute the scattering amplitude in Wilson Chiral Perturbation Theory for two flavors. The
lattice spacing effects due to the explicit chiral symmetry breaking are kept through O($a^2$), and we
consider the regime where the quark mass $m$ is of order $a^2 \Lambda_{\rm QCD}^3$. Analytic expressions for the scattering
lengths in different isospin channels are given. As a result of the O($a^2$) terms the scattering lengths
do not vanish in the chiral limit. Moreover, additional chiral logarithms proportional to $a^2 \ln M_{\pi}^2$ are
present in the one-loop results. These contributions can obscure the continuum chiral logarithms and
the determination of low-energy constants from numerical lattice simulations.
}

\FullConference{The XXVI International Symposium on Lattice Field Theory \\
		 July 14 - 19, 2008\\
		 Williamsburg, Virginia, USA}

\begin{document}

\section{Introduction}
Lattice simulations with light Wilson quarks have been made possible by significant algorihmic advances and increased computer power. For example, the PACS-CS collaboration recently reported results for pion masses as small as 156 MeV \cite{Aoki:2008sm}, a value that would have been thought impossible to reach a few years ago.

As much as one appreciates this developement, it also raises the question about the size of the lattice spacing artifacts. These become more and more pronounced as one decreases the quark mass keeping the lattice spacing fixed. In particular, one expects a sizable impact of the lattice artifacts in the regime where the quark mass $m$ is of the order $a^2\Lambda_{\rm QCD}^3$. In this case  the explicit breaking of chiral symmetry due to the quark mass and due to the Wilson term are of comparable size \cite{Sharpe:1998xm}. 

The appropriate tool to study this regime is Wilson chiral perturbation theory (WChPT) \cite{Sharpe:1998xm,Rupak:2002sm}. 
Here we report some of our recently obtained results for pion scattering in WChPT \cite{Aoki:2008gy,DABied}. We computed the scattering amplitude and the scattering lengths for all three isospin channels to one loop order in the chiral expansion. As a result of the O($a^2$) terms the scattering lengths
do not vanish in the chiral limit, and chiral logarithms proportional to $a^2 \ln M_{\pi}^2$ are present in the one-loop results. These contributions can obscure the continuum chiral logarithms and compromise the determination of Gasser-Leutwyler coefficients from numerical lattice simulations. Using the expression presented here should help in that respect.

\section{Setup}
We consider WChPT for $N_f=2$ degenerate quark flavors. In this case the chiral effective Lagrangian including the $p^2,m$ and $a^2$ term reads \cite{Bar:2003mh,Aoki:2003yv}
\begin{eqnarray}\label{L2}
{\cal L}_{\rm LO} & =& \frac{f^{2}}{4} \langle \partial_{\mu}\Sigma \partial_{\mu}\Sigma^{\dagger}\rangle  - \frac{f^{2}B}{2} {m} \langle \Sigma + \Sigma^{\dagger}\rangle + \frac{f^2}{16}\,c_2{a}^2\langle{\Sigma+\Sigma^\dagger}\rangle^2,
\end{eqnarray}
where $\Sigma$ contains the pion fields in the usual way and $\langle\ldots\rangle$ denotes the trace over the flavor indices. $f,B$ are the familiar LO low-energy coefficients (LEC) of continuum ChPT, and $c_2$ is an additional coefficient associated with non-zero lattice spacing effects.  

We keep the O$(a^2)$ correction in ${\cal L}_{\rm LO}$ and therefore promote it to a LO term in the chiral expansion. This is justified and necessary in the regime where $m\sim a^2\Lambda_{\rm QCD}^3$, the so-called {\em large cut-off effects} (LCE) regime. In the 
{\em generic small quark mass} (GSM) regime \cite{Sharpe:2004ps,Sharpe:2004ny}, one  assumes $m\sim a\Lambda_{\rm QCD}^2$. Results in this regime are easily obtained by properly expanding the results for the LCE regime.

The sign of $c_2$ determines the phase diagram of the theory \cite{Sharpe:1998xm}. If $c_2>0$ there exists an Aoki phase where parity and flavor are spontaneously broken \cite{Aoki:1983qi}. The charged pions are massless in this phase due to the spontaneous breaking of the flavor symmetry. Outside this phase the pion mass is given by
\bqry
\lbl{Mpitree}
M_0^2 & =& 2Bm - 2c_2a^2\,,
\eqry
and it vanishes at $m = c_2a^2/B$. If $c_2<0$ there exists a first order phase transition. 
All three pions are massive for all quark masses, and the pion mass assumes its minimal value   
\bqry
\label{Mpimin}
M_{0,{\rm min}}^2 & =& 2|c_2|a^2\,,
\eqry
at $m=0$.

\section{Pion scattering at tree-level} 

The scattering amplitude is  straightforwardly calculated  from the Lagrangian \pref{L2}.   At  tree-level we obtain 
\bqry
\lbl{amplitudetree}
A(s,t,u) & = & \frac{1}{f^2}(s - \mpit - \tct)\,,
\eqry
in terms of the three standard Mandelstam variables.  

Setting the lattice spacing to zero we recover the familiar result of continuum ChPT \cite{Gasser:1983yg}. At non-zero lattice spacing the amplitude receives a correction proportional to $c_2$. This modification also affects the scattering lengths 
$a_0^I$ for definite isospin $I=0$ and $I=2$:
\bqry
a_0^0 & =&\phantom{-} \frac{7}{32\pi f^2} \left(\mpit - \frac{5}{7}\, \tct\right)\,,\qquad
a_0^2 \, = \,  - \frac{1}{16\pi f^2} \Bigg(\mpit + \tct \Bigg)\,.\lbl{treescatlen}
\eqry
For $a=0$ we recover the well-known results first obtained by Weinberg \cite{Weinberg:1966kf}. For non-zero lattice spacings, however, the continuum results are modified in such a way that the scattering lengths no longer vanish in the chiral limit. Instead, they assume non-zero values of O($a^2$). Consequenty, the ratio $a_0^I/M_{0}^2$ is not a constant but has the functional form
\bqry
\lbl{treeadivmpi}
\frac{a_0^I}{M_{0}^2} & =& \frac{A_{00}^I}{M_{0}^2} + A^I_{10}\,,
\eqry
with $A^I_{10}$ being a constant and $A^I_{00}$ being of order $a^2$. Hence, the ratio $a_0^I/M_{0}^2$ diverges in the chiral limit. 
This divergence was first predicted by Kawamoto and Smit \cite{Kawamoto:1981hw}. However, note that the coefficient $A^I_{00}$ is of O($a^2$) rather than of O($a$). This holds even for unimproved Wilson fermions, in contrast to what has been expected earlier \cite{Sharpe:1992pp,Gupta:1993rn}.

The divergence in the chiral limit
will only be present if $c_2>0$, because only in this case can the pion indeed become massless. For $c_2<0$ the pion mass cannot be smaller than the minimal value in eq.\ \pref{Mpimin}, resulting in the  minimal values 
\bqry
\lbl{a0min}
a_{0,{\rm min}}^0 & =& \frac{12}{32\pi f^2} 2|c_2|a^2\,,\qquad
a_{0,{\rm min}}^2 \, = \, 0\,.
\eqry
for the scattering lengths.

%=================================================
%: figure 1
%=============
\begin{figure*}[t]
\begin{center}
 \includegraphics[scale=0.8]{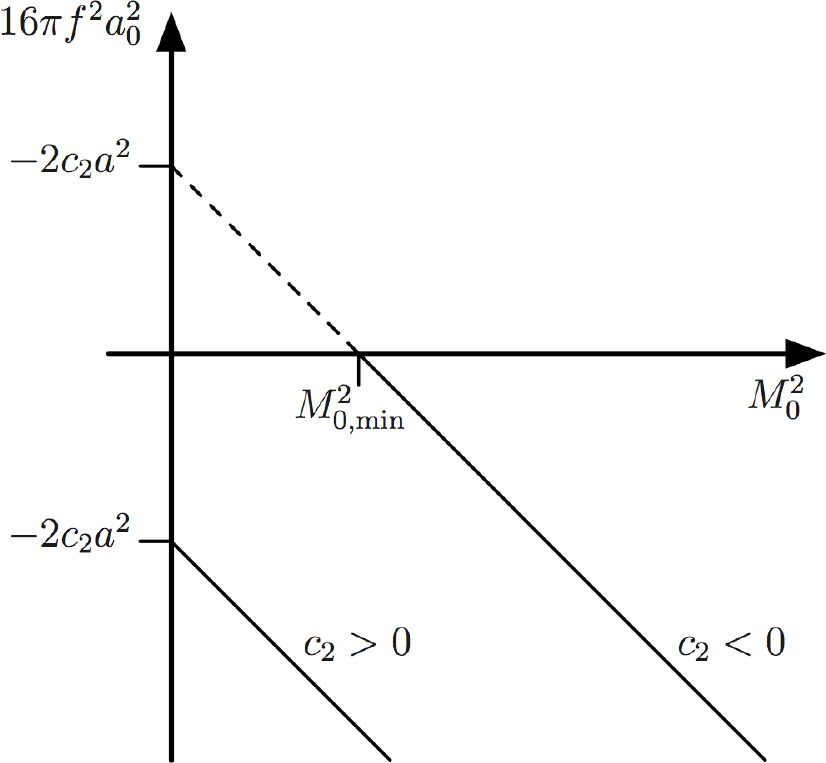}
 ~\hfill~
\includegraphics[scale=0.8]{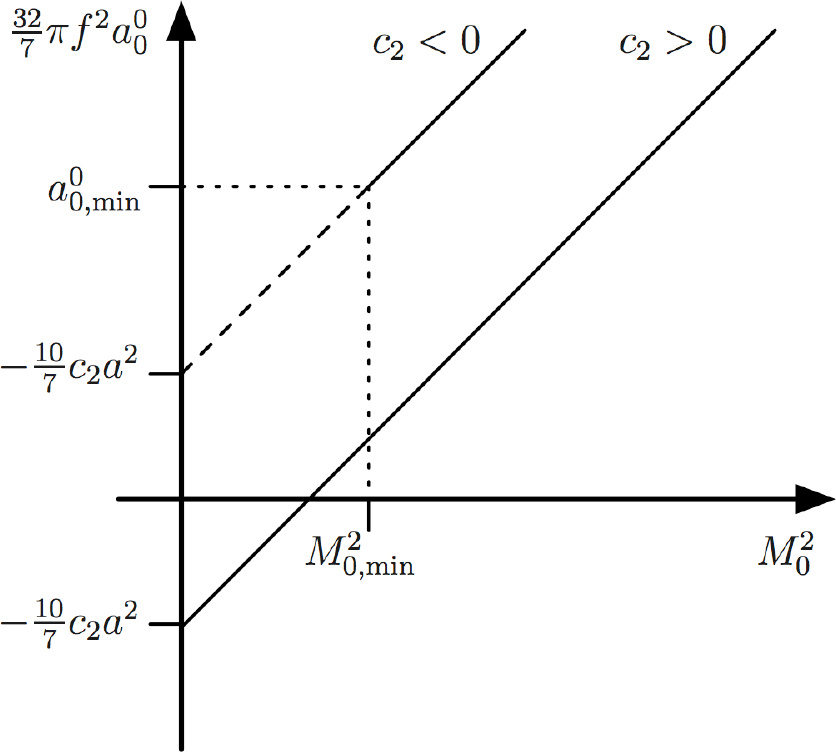}
 \end{center}
%\vspace*{0.3cm}   
\caption{Sketch of the scattering lengths as a function of the pion mass at non-zero lattice spacing. The left panel shows $16\pi f^2a_0^2$ as a function of the tree level pion mass $M_0^2$. For $c_2<0$ the pion mass cannot be smaller than $M^2_{0,{\rm min}}$. Nevertheless, extrapolating to the massless point the scattering length assumes the value $-2c_2a^2$, as indicated by the dashed line. For $c_2>0$ the pion mass can be taken zero. At this mass the scattering length also assumes the value $-2c_2a^2$, now with the opposite sign. The right panel shows the analogous sketch for the $I=0$ scattering length.
 }
\label{fig:1}      
\end{figure*}
%=================================================

Figure \ref{fig:1} sketches the pion mass dependence of the scattering lengths for the two possible signs of $c_2$. It seems feasible that measurements of  the scattering lengths will allow determinations of  $c_2$. Extrapolating the data for $a_0^2$ to the chiral limit  one may directly read off  $c_2$ as the value at vanishing pion mass. A practical advantage is that the calculation of $a_0^2$ does not involve disconnected diagrams which introduce large statistical uncertainties.

The $I=1$ channel is somewhat special. The scattering amplitude for this isospin channel is given by  $A(t,s,u)-A(u,t,s)$. The $c_2$ contribution drops out in this difference and the scattering length $a_1^1$ is just as in continuum ChPT \cite{Gasser:1983yg}. This suggests that the scaling violations in $a_1^1$ are small.

\section{Pion scattering at one-loop}

\subsection{Power counting in the LCE regime}
The power counting in WChPT is slightly non-trivial if we take the O($a^2$) term at LO. ${\cal L}_{\rm LO}$ consists of the terms of O($p^2,m,a^2$). In order to renormalize the divergencies of the loop diagrams one needs higher order counterterms in the chiral Lagrangian. These are, besides the standard ones of continuum ChPT \cite{Gasser:1983yg,Gasser:1984gg}, the terms of order $p^2a^2, ma^2,a^4$. In addition, terms of order $p^2a, ma,a^3$ also exist and are formally of lower order. Consequently, these should also be included in the chiral expansion. Hence, one-loop calculations in the LCE regime require the following terms in the chiral Lagrangian:\\
\vspace{-0.8cm}
\bqry
\lbl{PcountingLCE}
\begin{array}{rcl}
{\rm LO:}& \quad & p^2,\,m,\,a^2\\
{\rm NLO:}& \quad & p^2a,\,ma,\,a^3\\
{\rm NNLO:}& \quad & p^4,\,p^2m,\,m^2,\,p^2a^2,\, ma^2,\,a^4
\end{array}
\eqry
The standard NLO terms of continuum ChPT appear here at NNLO, a consequence of the fact that $m\sim a^2\Lambda_{\rm QCD}^3$ in the LCE regime.

\subsection{The pion mass to one loop}

The modifications due to extra chiral logarithms are already seen in the one-loop result for the pion mass. In terms of the tree-level mass $M_0^2$ we find \cite{Aoki:2003yv}
\bqry
\lbl{pionmassoneloop}
 M_{\pi}^2&=&M^2_0\left[1+\frac{1}{32\pi^2}\frac{M^2_0}{f^2}\ln\left(\frac{M^2_0}{\Lambda_3^2}\right)+\frac{5}{32\pi^2}\frac{2c_2a^2}{f^2}\ln\left(\frac{M^2_0}{\Xi_3^2}\right) +k_1\frac{W_0a}{f^2}\right]\nonumber \\
 &&\,+\,k_3\frac{2c_2W_0a^3}{f^2}+k_4\frac{(2c_2a^2)^2}{f^2}\,.
\eqry
The coefficients $\Lambda_3^2,\Xi_3^2$ and $k_1,k_3,k_4$ are (combinations of) unknown LECs. Eq.\ \pref{pionmassoneloop} reproduces the well-known result in the naive continuum limit, but is modified for non-zero $a$, in particular by a chiral logarithm proportional to $c_2a^2$. Moreover, the coefficient in front of the chiral log is much larger than the coefficient in front of the continuum chiral log. Consequently, already small values of $c_2a^2$ may dilute the continuum log and the characteristic curvature one usually looks for in numerical lattice data might be significantly diminished.

We remark that the ETM collaboration has measured a negative value for $c_2$ in their twisted mass simulations. The data  \cite{Urbach:2007rt} for $a \approx 0.086$ fm and $M_{\pi^{\pm}}\approx 300$ MeV results in $-2c_2a^2 \approx (185$ MeV$)^2$, determined from the pion mass splitting. Such a value completely suppresses the chiral log for pion masses around 400 MeV, a value not unusual in present day lattice simulations.

\subsection{The $I=2$ scattering length at one loop}

%=================================================
%: figure 2
%=============
\begin{figure*}[t]
\begin{center}
\includegraphics[scale=0.6]{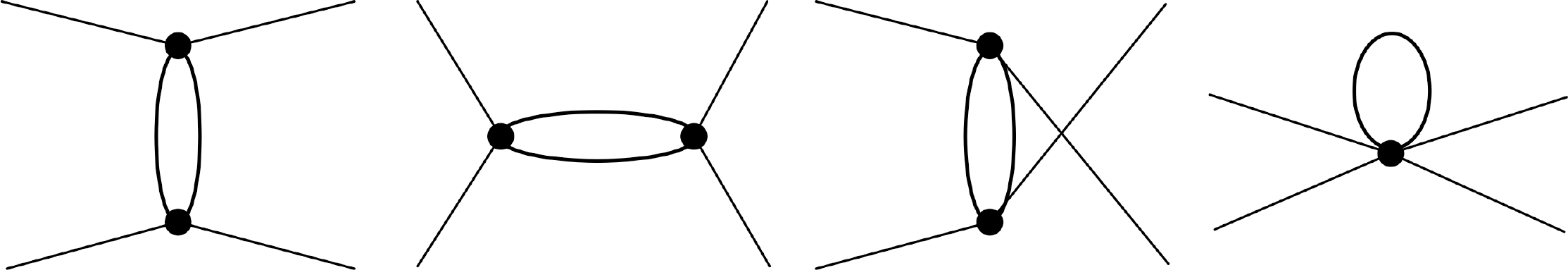}
 \end{center}
%\vspace*{0.3cm}   
\caption{One-loop diagramms contributing to the four-point function.
The diagrams with one or two vertices stemming from the O($a^2$) term in the chiral Lagrangian will give rise to chiral logs proportional $ c_2a^2M_{0}^2\ln M_{0}^2$ and $(c_2a^2)^2\ln M_{0}^2$.
 }
\label{fig:3}      
\end{figure*}
%=================================================

The four diagrams in figure \ref{fig:3} contribute to the four-point function at one loop. The vertices in these diagrams can be either a vertex also present in continuum ChPT or the vertex proportional to $c_2a^2$ stemming from the O($a^2$) term in Eq.\ \pref{L2}. The latter give rise to additional chiral logarithms $c_2a^2M_0^2\ln M_0^2 $  and $(c_2a^2)^2\ln M_0^2$ in the scattering amplitude and the scattering lengths. 

As an example we give the one-loop result for the $I=2$ scattering length (for the other isospin channels see Ref.\  \cite{Aoki:2008gy}):
\begin{eqnarray}
a_0^2&=&-\frac{M_{\pi}^2}{16\pi
 f^2}\left(\kappa_{21} + \frac{M_{\pi}^2}{16\pi^2f^2}\left\{\frac{7}{2} \ln\frac{M_{\pi}^2}{\mu^2} - \frac{4}{3}l^{\rm I=2}_{\pi\pi}\right\}
+\frac{2c_2a^2}{16\pi^2f^2}\left\{3\ln\frac{M_{\pi}^2}{\mu^2} \right\} \right)
 \nonumber\\
 &&- \frac{2c_2a^2}{16\pi f^2}\left(\kappa_{22} + \frac{2c_2a^2}{16\pi^2 f^2}\left\{\frac{11}{2}\ln\frac{M_{\pi}^2}{\mu^2}\right\} \right)\label{resulta2}
\end{eqnarray}
The parameters $\kappa_{2j}$ comprise all the analytic terms through O($a^2$). Therefore, these are constants for fixed $a$ and serve as fitting parameters in the analysis of numerical data. As a short hand notation we introduced
\bqry
l^{\rm I=2}_{\pi\pi} \, =\, l_1 + 2l_2 -\frac{3}{8} l_3 + \frac{3}{8}\,\lbl{LECcomb}
\eqry
for the combination of LECs entering the scattering lengths $a_0^2$.\footnote{We emphasize that  \pref{resulta2} is expressed  in terms of the decay constant in the chiral limit, and not in terms of $f_{\pi}$. If the latter had been used the LEC $l_4$ \cite{Gasser:1983yg} would also appear in \pref{LECcomb}. }  

As expected, \pref{resulta2} reproduces the continuum result for $a\rightarrow0$. For non-zero $a$, however, it differs significantly from the corresponding one in continuum ChPT. 
The scattering length does not vanish in the chiral limit, but rather seems to diverge as $a^4\ln M_{\pi}^2$. 
However, if $c_2 <0$ no divergence is present since the pion mass is bounded by the minimal value in eq.\ \pref{Mpimin}. If $c_2>0$  massless pions are in principle possible, but for small pion masses the chiral expansion eventually breaks down when $c_2a^2\ln M_{\pi}^2$ becomes of order unity. In that case higher order terms leading to powers $(c_2a^2\ln M_{\pi}^2)^n, n=2,3,\ldots$ become relevant  and a summation of all these terms is necessary.\footnote{This resummation can presumably be done along the lines in Ref.\ \cite{Aoki:2003yv}, where a resummed pion mass formula has been derived.}
This should be kept in mind if one keeps lowering the quark mass at a fixed lattice spacing: Eventually the lattice spacing corrections will be enhanced by the unphysical divergence $a^4\ln M_{\pi}^2$.
 
Result \pref{resulta2} contains five unknown parameters: the continuum parameters $f$ and $l_{\pi\pi}^{\rm I=2}$ as well as $\kappa_{21}, \kappa_{22},c_2$, i.e.\ three more than the continuum result.
Consequently, the pion mass dependence of the scattering length on the lattice can be very different in contrast to what one may expect from continuum ChPT. 
Attempts to fit lattice data using expressions from continuum ChPT may easily fail. 

The ratio $a_0^{2}/M_{\pi}^2$ has the functional form
\bqry
\lbl{loopadivmpi}
\frac{a_0^{2}}{M_{\pi}^2} & =& \frac{A_{00}}{M_{\pi}^2} + A_{10} + A_{20}M_{\pi}^2 + A_{30}M_{\pi}^2\ln M_{\pi}^2 + A_{40}\ln M_{\pi}^2+ \tilde{A}_{40}\frac{\ln M_{\pi}^2}{ M_{\pi}^2}\,.
\eqry
The first two terms on the right hand side correspond to the tree level result in eq.\ \pref{treeadivmpi}.  The constants $A_{00} - A_{40}$ represent the five independent fit parameters, while $\tilde{A}_{40}$ is not independent. The first three terms in \pref{loopadivmpi} have already been used in analyzing numerical lattice data \cite{Yamazaki:2003za}, but the data  could not be fitted well.  It might be interesting to repeat the analysis with the full result in \pref{loopadivmpi}, even though the data was obtained for heavy pion masses between 500 MeV and 1.1 GeV, and ChPT is not expected to be applicable.

\section{Conclusions}

For pion scattering with Wilson fermions we essentially find two modifications due to a non-zero lattice spacing. First, the  $I=0,2$ scattering lengths do not vanish in the chiral limit, and second, additional chiral logarithms proportional to $a^2$ appear in the one-loop results for these quantities.   Ignoring these modifications and using the results of continuum ChPT in the analysis of numerical lattice data is potentially dangerous, depending on the size of these extra contributions. In particular, the extra contributions proportional to the lattice spacing become more and more relevant the smaller the quark mass is. This should be kept in mind when one pushes to the physical point at non-zero lattice spacing.

We conclude with a comment on the coefficient $c_2$, which determines the phase diagram of the theory.  The $I=2$ scattering length may provide a handle to obtain an estimate of $c_2$. A practical advantage is that the calculation of $a_0^2$ does not involve disconnected diagrams which introduce large statistical uncertainties.
We expect that at least the sign of $c_2$ is easily accessible, and this is what matters for the phase diagram.  

\section*{Acknowledgments}

This work is supported in part by the Grants-in-Aid for
Scientific Research from the Japanese Ministry of Education,
Culture, Sports, Science and Technology
(No.\ 20340047) and by the Deutsche Forschungsgemeinschaft (SFB/TR 09).
B. B.\  acknowledges financial support from the {\em Cusanuswerk}.

\end{document}